# Harnessing van der Waals CrPS$_4$ and Surface Oxides for non-monotonic pre-set field induced Exchange Bias in Fe$_3$GeTe$_2$

Aravind Puthirath Balan[1‡*], Aditya Kumar[1‡], Tanja Scholz[2], Zhongchong Lin[3], Aga Shahee[1], Shuai Fu[4], Thibaud Denneulin[5], Joseph Vas[5], András Kovács[5], Rafal E. Dunin-Borkowski[5], Hai I. Wang[4], Jinbo Yang[3], Bettina Lotsch[2], Ulrich Nowak[6], Mathias Kläui[1,7*]

[1]Institute of Physics, Johannes Gutenberg University Mainz, Staudinger Weg 7, 55128 Mainz, Germany.

[2]Max Planck Institute for Solid State Research, Heisenbergstraße 1, 70569 Stuttgart, Germany.

[3]State Key Laboratory for Artificial Microstructure and Mesoscopic Physics, School of Physics, Peking University, Beijing 100871, China.

[4]Max Planck Institute for Polymer Research Mainz, Ackermannweg 10, 55128 Mainz, Germany

[5]Ernst Ruska-Centre for Microscopy and Spectroscopy with Electrons and Peter Grünberg Institute, Forschungszentrum Jülich, 52425 Jülich, Germany

[6]Department of Physics, University of Konstanz, Universitaetsstrasse 10, 78464 Konstanz, Germany

[7]Centre for Quantum Spintronics, Department of Physics, Norwegian University of Science and Technology, 7491 Trondheim, Norway.

[‡]These authors contributed equally
* Correspondence to aravindputhirath@uni-mainz.de, klaeui@uni-mainz.de

**Abstract**

Two-dimensional van der Waals (vdW) heterostructures are an attractive platform for studying exchange bias due to their defect free and atomically flat interfaces. Chromium thiophosphate ($CrPS_4$), an antiferromagnetic material, possesses uncompensated magnetic spins in a single layer, rendering it a promising candidate for exploring exchange bias phenomena. Recent findings have highlighted that naturally oxidized vdW ferromagnetic $Fe_3GeTe_2$ exhibits exchange bias, attributed to the antiferromagnetic coupling of its ultrathin surface oxide layer (O-FGT) with the underlying unoxidized $Fe_3GeTe_2$. Anomalous Hall measurements are employed to scrutinize the exchange bias within the $CrPS_4$/(O-FGT)/$Fe_3GeTe_2$ heterostructure. This analysis takes into account the contributions from both the perfectly uncompensated interfacial $CrPS_4$ layer and the interfacial oxide layer. Intriguingly, a distinct and non-monotonic exchange bias trend is observed as a function of temperature below 140 K. The occurrence of exchange bias induced by a 'pre-set field' implies that the prevailing phase in the polycrystalline surface oxide is ferrimagnetic $Fe_3O_4$. Moreover, the exchange bias induced by the ferrimagnetic $Fe_3O_4$ is significantly modulated by the presence of the van der Waals antiferromagnetic $CrPS_4$ layer, forming a heterostructure, along with additional iron oxide phases within the oxide layer. These findings underscore the intricate and complex nature of exchange bias in van der Waals heterostructures, highlighting their potential for tailored manipulation and control.

**KEYWORDS:** vdW magnetic materials, exchange bias, vdW heterostructures, $Fe_3GeTe_2$, $CrPS_4$

The phenomenon of exchange bias (EB) arises when a ferromagnetic (FM) material and an antiferromagnetic (AFM) material are brought into contact, resulting in the development of a unidirectional magnetic anisotropy at their interface[1]. This introduced anisotropy causes a shift in the magnetic hysteresis loop of the FM along the field axis. EB has significant implications in various device technologies, such as spin valves, which find applications in high-density magnetic storage and sensor systems[2]. The underlying mechanism behind EB is generally attributed to the exchange coupling between uncompensated spins within the AFM and the FM moments across the interface[3,4]. However, in thin-film heterostructures comprising non-van der Waals solids, the presence of imperfections like interdiffusion, step edges, grain boundaries, interfacial stress, and roughness can compromise the ideal interface quality[5].

The recent emergence of van der Waals (vdW) magnetic materials exhibiting atomically smooth surfaces has revitalized interest in understanding the microscopic origins of EB[6]. With their inherent layered structure, vdW materials offer the potential to form heterostructures with nearly ideal interfaces, making them particularly promising for investigating interface-related effects, especially pertaining to EB[7]. Recent reports have unveiled the occurrence of EB in vdW AFM/FM heterostructures. For instance, an EB of approximately 50 mT was identified in a $CrCl_3/Fe_3GeTe_2$ heterostructure at 2.5 K[8]. Subsequent investigations involved various vdW AFM layers ($MnPS_3$, $MnPSe_3$, $FePS_3$) atop metallic $Fe_3GeTe_2$ or $Fe_5GeTe_2$[9,10]. A substantial EB was also observed in a heterostructure composed of a vdW FM insulator and an AFM topological insulator ($CrI_3/MnBi_2Te_4$) with partial coverage[11]. Another notable development is the observation of EB in naturally occurring superlattice structures of $MnBi_2Te_4(Bi_2Te_3)n$ (n = 1, 2), attributed to exchange coupling between co-existing ferromagnetic and antiferromagnetic constituents in the ground state[12].

Among vdW antiferromagnets, $CrPS_4$ stands out due to its fully uncompensated layered structure, enabling ideal coupling at the interface within vdW AFM/FM heterostructures[13,14,15].

Moreover, $Fe_3GeTe_2$ (FGT) is an itinerant vdW ferromagnet boasting a substantial Curie temperature of 220 K in bulk[16]. Recent investigations have highlighted that an air-exposed $Fe_3GeTe_2$ crystal naturally forms a thin antiferromagnetic surface oxide layer (O-FGT), resulting in EB in the underlying unoxidized crystal[17]. Analytical techniques, including energy dispersive spectroscopy mapping, X-ray absorption spectroscopy, and electron energy loss spectroscopy, have indicated a polycrystalline nature for the oxide layer, suggesting the presence of diverse phases such as FeO, $Fe_2O_3$, and $Fe_3O_4$ corresponding to varying $Fe^{2+}$ and $Fe^{3+}$ oxidation states[18,19]. This surface oxide has also been found to enhance EB in $CrOCl/Fe_3GeTe_2$ heterostructures[20]. Notably, significant EB has been reported in naturally oxidized and restacked liquid exfoliated $Fe_3GeTe_2$ nanoflakes[21]. Nevertheless, investigations elucidating the precise magnetic characteristics of the formed oxide layer and its role in introducing EB remain lacking.

This study focuses on the characterization of EB properties within vdW heterostructures, specifically $CrPS_4/(O\text{-}FGT)/Fe_3GeTe_2$ and $CrPS_4/Fe_3GeTe_2$, using magneto-transport measurements. Our findings unveil the intricate magnetic signatures underlying the observed robust EB in $CrPS_4/(O\text{-}FGT)/Fe_3GeTe_2$, which primarily stems from the dominance of ferrimagnetic magnetite within the surface oxide at the interface. Furthermore, we ascertain that the EB attributed to magnetite is influenced by the presence of vdW $CrPS_4$ in a heterostructure and the coexistence of a second antiferromagnetic iron oxide phase in the polycrystalline surface oxide. Through a comprehensive analysis of both the complex EB in $CrPS_4/(O\text{-}FGT)/Fe_3GeTe_2$ and the intrinsic EB in $CrPS_4/Fe_3GeTe_2$ featuring a pristine interface, this study provides insights into tailoring EB in vdW heterostructures.

**Results and Discussions**

Single crystals of Fe$_3$GeTe$_2$ (FGT) and CrPS$_4$ are grown using chemical vapor transport following previous reports[22, 23]. The quality of synthesized crystals is examined by Raman spectroscopy (please refer to *Figure S1 in Supporting Information* under *Section 1*). All the modes obtained for both Fe$_3$GeTe$_2$ and CrPS$_4$ bulk crystals match well with predicted Raman modes [24,25]. To explore the impact of O-FGT layer on EB, two heterostructures are fabricated: CrPS$_4$/(O-FGT)/Fe$_3$GeTe$_2$ and CrPS$_4$/Fe$_3$GeTe$_2$. In the former, a surface oxide layer coats the surface of the Fe$_3$GeTe$_2$ flake at the interface, while the latter features a pristine Fe$_3$GeTe$_2$ surface. The surface oxide formation is confirmed using element-sensitive cross-sectional transmission electron microscopy measurements (please refer to *Figure S2a-h* in *Supporting Information* under *Section 2* for details) of a CrPS$_4$/(O-FGT)/Fe$_3$GeTe$_2$ heterostructure. The thickness of the Fe$_3$GeTe$_2$ and CrPS$_4$ layers of the exposed heterostructure is determined to be 30 nm and 100 nm, respectively, using atomic force microscope (for details please refer to *Figure S3* in *Supporting Information* under *Section 3*). CrPS$_4$/(O-FGT)/Fe$_3$GeTe$_2$ heterostructures are fabricated on to a prepatterned anomalous Hall contacts and were subjected to magneto-transport anomalous Hall voltage ($V_{xy}$) measurements. Please refer to the methods section for more details about the device fabrication and anomalous Hall voltage measurements. The results obtained are summarised in *Figure 1*. Exchange bias field ($H_{EB}$) is determined from $V_{xy}$ measured as a function of sweeping magnetic field as $H_{EB} = \left(\frac{H_C^+ + H_C^-}{2}\right)$, where $H_C^+$ and $H_C^-$ represent the coercivities at positive and negative fields, respectively.

Initially, we used a magnetic field of +8 T to cool down the CrPS$_4$/(O-FGT)/Fe$_3$GeTe$_2$ sample from room temperature to the desired temperatures. Then, we determined the corresponding $H_{EB}$ at these temperatures by analyzing $V_{xy}$ measured as a function of sweeping magnetic field. The process is repeated for several temperature set points from 5 K up to 150 K and the results are summarized in *Figure 1a*. $H_{EB}$ is calculated for all of them and plotted as a function of temperature (*Figure 1d*). It is generally expected that the magnitude of $H_{EB}$ is maximum at low temperatures and gradually decreases as we go to higher temperatures and disappears above the blocking temperature ($T_B$) of the antiferromagnet. However, the $H_{EB}$ here follows a complex non-monotonic trend with the temperature. As indicated by the curve in green circles in *Figure 1d*, the observed trend consists of three distinct regions (indicated by three different colors) characterized by two local minima at ~20 K and ~70 K, respectively. The robust EB observed up to 140 K can be attributed to the surface oxide formation on $Fe_3GeTe_2$ at the interface as previously reported[17]. However, the complex trend with temperature that we observe has not been studied and reported before. To understand the mechanisms on the low temperature regime, *i.e*., 5 K to 36 K, one needs to analyze the EB on an h-BN capped $CrPS_4$/$Fe_3GeTe_2$ device with a pristine interface without any oxide formation (please refer to the *methods* section for details of device fabrication).

To analyze the EB of the device with a pristine interface (see *Figure 2a* for the optical image), the device is field-cooled from room temperature to temperatures below the Néel temperature of $CrPS_4$ ($T_N = 36\ K$) with an out of the plane field of + 8 T. After the field cooling, $V_{xy}$ is measured as a function of sweeping magnetic field at various temperatures from 5 K to 50 K and are summarized in *Figure 2b*. The directions of cooling field ($H_{FC}$) and EB observed are indicated by black and red arrows respectively. $H_{EB}$ is calculated and plotted as a function of temperature (*Figure 2c*). At 5 K, the magnitude of $H_{EB}$ for a $CrPS_4$/$Fe_3GeTe_2$ device with a pristine interface is found to be ~ 12.5 mT ($\pm 1\ mT$) which gradually decreases as the temperature is increased and found to be negligible ($< 2\ mT$) above 20 K which is the $T_B$ of $CrPS_4$ as already reported previously[22]. We also observed a switching of the polarity of EB from negative to positive at 15 K close to the $T_B$ similar to the observations by *S. Ding et al*., where they observed switching EB behavior in pristine $CrPS_4$/FeNi interface[22].

Interestingly, the observed $T_B$ for a CrPS$_4$/Fe$_3$GeTe$_2$ system is found to be 20 K, significantly lower than the $T_N$ of CrPS$_4$. An estimation of the interfacial exchange coupling energy between CrPS$_4$ and Fe$_3$GeTe$_2$, calculated using a simplistic model provided by *D Mauri et al.*[26], is comparable to the weaker interlayer antiferromagnetic exchange coupling (0.16 meV) in CrPS$_4$[15]. This indicates that the exchange field from Fe$_3$GeTe$_2$ could be as strong as the spin-flip field in CrPS$_4$ and could potentially be even stronger as the temperature approaches $T_N$ of CrPS$_4$ and the spin-flip field decreases[15]. In case of the pristine CrPS$_4$/Fe$_3$GeTe$_2$ system, at temperatures above 20 K, the exchange field may exceed the spin-flip field. Consequently, CrPS$_4$ would flip with Fe$_3$GeTe$_2$, resulting in zero EB, thus explaining the lower $T_B$ observed.

As shown in *Figure 2d*, the variation of $H_{EB}$ for a CrPS$_4$/(O-FGT)/Fe$_3$GeTe$_2$ (orange curve) follows a similar trend to that observed for a pristine CrPS$_4$/Fe$_3$GeTe$_2$ interface (blue curve), albeit with an enhanced magnitude. The enhancement in the magnitude of $H_{EB}$ observed here is analogous to what has been observed previously in CrOCl/(O-FGT)/Fe$_3$GeTe$_2$ heterostructures[20]. The presence of the O-FGT layer enhances the overall interface coupling strength, thereby increasing the $H_{EB}$. However, a comparatively smaller uncompensated moment in the O-FGT layer couples weakly with the interfacial CrPS$_4$ layer. This may result in CrPS$_4$ retaining its antiferromagnetic state even above the observed $T_B$ and can still induce EB up to $T_N$. Consequently, the magnitude of $H_{EB}$ in CrPS$_4$/(O-FGT)/Fe$_3$GeTe$_2$ increases from the minima at 15 K corresponding to the maximum positive EB in the pristine device up to $T_N$ (CrPS$_4$) = 36 K (refer to *Figure 2d*) where the positive EB from CrPS$_4$ disappears. The second minimum at 70 K (*Figure 1d*) further suggests that two different phases in the surface oxide contribute to EB, which is consistent with previous reports[18,19]. However, determining the exact nature of oxide phases is challenging considering the polycrystalline nature of the surface oxide. Our investigation of EB in CrPS$_4$/(O-FGT)/Fe$_3$GeTe$_2$ helps us unravel the nature of two distinct phases of oxides in O-FGT through their distinct EB fingerprints.

Surprisingly, unlike the CrPS$_4$/ Fe$_3$GeTe$_2$ with a pristine interface, the CrPS$_4$/(O-FGT)/Fe$_3$GeTe$_2$ device found to be generating EB even without field-cooling by merely applying a 'pre-set field' ($H_{PF}$), and the $H_{EB}$ due to $\pm 1$ T pre-set field follows a similar trend as obtained for field-cooling (*Figure 1d*). Please see the methods section for more information about $H_{PF}$ and how it is different from conventional field-cooling. The $V_{xy}$ measurements for a CrPS$_4$/(O-FGT)/Fe$_3$GeTe$_2$ corresponding to all the three regions for both $H_{PF} = +1\,T$, and $H_{PF} = -1\,T$ are summarized in *Figure 1b and 1c*, respectively. $H_{EB}$ is calculated and plotted as a function of temperature (*Figure 1d*). Such a peculiar $H_{PF}$ induced EB observed in CrPS$_4$/(O-FGT)/Fe$_3$GeTe$_2$ heterostructure in the temperature range of 5 K to 140 K suggests that the dominant phase in the O-FGT layer that contributes to EB could be ferrimagnetic Fe$_3$O$_4$ analogous to ferrimagnet generated EB reported previously[27,28]. The dominance of ferrimagnetic Fe$_3$O$_4$ in the surface oxide layer is suggested as its magnetic alignment in a particular direction by a $H_{PF}$ determines the direction of pinning and hence the direction of loop shift [28].

Material characterization techniques such as cross-sectional scanning transmission electron microscopy (STEM) – electron energy loss spectroscopy (EELS) and surface sensitive x-ray photoelectron spectroscopy (XPS) (see *Figure S4* and *Figure S5* in *Supporting Information* under *Section 4* and *Section 5* respectively) are performed to check possible phases of oxides in O-FGT. XPS 2p$_{3/2}$ and 2p$_{1/2}$ peaks suggest the possible oxide phases are Fe$_3$O$_4$, and Fe$_2$O$_3$, with both Fe$^{3+}$ and Fe$^{2+}$ valence states. However, the analysis of the Fe-L3/L2 intensity ratio in EELS, in comparison with ratios obtained from spectra of reference samples, clearly suggests the presence of Fe$_3$O$_4$[19]. The presence of Fe$_3$O$_4$ is further supported by the observation of a sudden increase in longitudinal resistance in an Fe$_3$GeTe$_2$ flake with an O-FGT layer at around 125 K, indicating a Verwey transition in $Fe_3O_4$ (see *Figure S6a* in *Supporting Information* under *Section 6*).

Importantly, the minimum field required to saturate the ferrimagnetic $Fe_3O_4$ in the O-FGT at a particular temperature can be considered as the critical pre-set field ($H_{CF}$) for the system. Therefore, a $H_{PF}$ greater than the $H_{CF}$ in the preferred direction can be applied to judiciously determine the direction of EB (please refer *Figure S7* in *Supporting Information* under *Section 10*). When the field is swept with a maximum value greater than $H_{CF}$ at a given temperature preceded by a $H_{PF} \geq H_{CF}$, there is no preferred pinning by ferrimagnetic $Fe_3O_4$, resulting in zero EB (refer to *Figure S8a-d* in *Supporting Information* under *Section 8*)[28]. In addition to $Fe_3O_4$, previous reports confirmed the presence of an antiferromagnetic FeO phase in the natural O-FGT[18,19]. This co-existing FeO phase may have a low $T_B$ of around 70 K, which is below the bulk $T_N = 198$ K [29]. The FeO phase is responsible for modulating the EB up to 70 K, where we observed a second minimum corresponding to $T_B$ of FeO (*Figure 1d*). Beyond 70 K, the EB is solely due to ferrimagnetic $Fe_3O_4$, which disappears at 140 K, the identified $T_B$.

Finally, the complex trend of $H_{EB}$ with respect to the temperature can be understood based on an intuitive model of the contributions of $CrPS_4$ as well as the two iron oxide phases in the O-FGT layer. The intuitive model is schematically represented in *Figure 3*. The various magnetic layers are colour coded. The grey and yellow layers are the bottom and top layers of $CrPS_4$ and $Fe_3GeTe_2$ on either side of the vdW gap, whereas the O-FGT consists of effectively two layers: a thicker top layer mainly composed of $Fe_3O_4$, as suggested by cross-sectional STEM-EELS measurements (for details, please refer to *Figure S4* in *Supporting Information* under *Section 4*), and a thinner bottom layer composed of FeO, corresponding to the low valence state of Fe. The presence of two types of oxides layers is also evident from the different contrasts observed in the cross-sectional STEM image of the $CrPS_4$/(O-FGT) $Fe_3GeTe_2$ interface (*Figure 3a* inset). In the case of the first device with a pristine interface, only a field-cooling below the $T_N$ of CrPS$_4$ (36 K) can induce EB (*Figure 3a*), whereas, it remains unbiased for a $H_{PF}$ (for details, please refer to *Figure S9* in *Supporting Information* under *Section 9*). However, the reason for the observed positive EB close to the $T_B$ is still not understood properly. In the case of the second device with the O-FGT, however, three cases have to be considered as evident from *Figure 1d*: (i) T < 36 K (*Figure 3b*), where all the three constituents are magnetically ordered, (ii) 36 K < T < 70 K (*Figure 3c*), where CrPS$_4$ is paramagnetic, while both oxide phases are magnetically ordered, and (iii) 70 K < T < 140 K (*Figure 3d*), where only Fe$_3$O$_4$ is magnetically ordered, and contributes to EB since $T_B$ of FeO is at 70 K.

Here, if we analyze the $H_{EB}$ vs. *T* (*Figure 1d*), it is apparent that the dominant contribution is due to ferrimagnetic Fe$_3$O$_4$ since we are observing a $H_{PF}$ induced EB (*blue* and *orange* curves) throughout the temperatures from 5 K to 140 K. This is further reinforced by the absence of any EB due to pristine CrPS$_4$ after applying $H_{PF}$ (for details, please refer to *Figure S9* in *Supporting Information* under *Section 9*). However, CrPS$_4$ and FeO plays important roles in modulating dominant EB due to Fe$_3$O$_4$ in the first two regions where they are magnetically ordered and their contributions to EB are evident. In the first case (*Figure 3b*), since all the magnetic layers are in their magnetically ordered state, the resulting EB is then the net effect of all three EB contributions. The first local minimum of about ~20 K is identified to be the $T_B$ of CrPS$_4$. It is worth noting in *Figure 1d* that below 20 K, the influence of CrPS$_4$ on $H_{EB}$ observed in a CrPS$_4$/(O-FGT)/Fe$_3$GeTe$_2$ device seems to be different for the measurements preceded by a conventional field-cooling (indicated by *green* curve) and preceded by a $H_{PF}$ (indicated by *blue* and *orange* curves). $H_{EB}$ is found to be large for 1 T preset-field more than what we obtained for even higher cooling-field of 8 T (for detailed analysis, please refer *Figure S10* in *Supporting Information* under *Section 10*). This suggests that in addition to modulation, a field-cooled CrPS$_4$ can have converse effects on the magnitude of $H_{EB}$, whereas CrPS$_4$ under $H_{PF}$ application does not. Detailed spin structure information at a domain level is necessary to find out the underlying reason for this contrasting behavior, which is however beyond the scope of this investigation. Additionally, above 20 K, the first CrPS$_4$ layer close to the interface is exchange coupled to the magnetization of Fe$_3$GeTe$_2$ layer at the bottom due to the strong effective exchange field of saturated Fe$_3$GeTe$_2$. The strength of this additional exchange coupled CrPS$_4$ layer is maximum at 20 K and gradually decreases up to 36 K, the $T_N$ of CrPS$_4$ and then disappears. This contributes to a slight increase of $|H_{EB}|$ from the minimum at 20 K to 36 K.

In the second case (*Figure 3c*), CrPS$_4$ is paramagnetic, and the net EB contribution could be the sum of the contributions of Fe$_3$O$_4$ and FeO, with the second local minimum at 70 K corresponding to the $T_B$ of FeO. Finally, above 70 K (*Figure 3d*), the EB contribution is only originating from ferrimagnetic Fe$_3$O$_4$ that acts directly as the source of pinning, however, through a thin and largely magnetically inactive spacer layer that mainly consists of FeO. This might weaken the longer distance exchange coupling, as evident by the significant training effect above 70 K (see *Figure S11* in *Supporting Information* under *Section 11*). We observe that the $H_{EB}$, exclusively induced by Fe$_3$O$_4$, is notably elevated at 80 K, exhibiting a comparative strength nearly equivalent to the $H_{EB}$ observed at the significantly lower temperature of 5 K. This sudden increase in strength above 70 K is reasonable taking into account that the major pinning source is the Fe$_3$O$_4$ layer, which acts directly on the Fe$_3$GeTe$_2$ without any magnetically active intermediate layers, in contrast to the case where FeO is active below 70 K.

**Conclusions**

In summary, we have successfully demonstrated exchange bias in CrPS$_4$/(O-FGT)/Fe$_3$GeTe$_2$ vdW heterostructures, where we observed a complex non-monotonic trend of exchange bias with temperature. The observed trend is apparent for both field-cooling as well as for a 'pre-set field'. The two local minima observed in the exchange bias trend with the temperature are identified to correspond to the $T_B$ of CrPS$_4$ and FeO. Saturating the ferrimagnetic Fe$_3$O$_4$ in the surface oxide by applying a field above a 'critical pre-set field' determines the direction of the unidirectional exchange anisotropy and hence the direction of loop shift, which then allows us to tune the exchange bias. An intuitive model is proposed to explain the complex non-monotonic trend of exchange bias fields as a function of temperature based on the different contributions. Our investigation identifies an alternative approach to tailor exchange bias in heterostructures by extrinsic methods such as oxidation combined with applied fields. We believe this investigation will motivate researchers to explore the exceptional tunability of magnetic properties in vdW heterostructures for spintronic applications. Given the large number of AFM vdW materials, our work provides insight into how surface oxide formation on the Fe$_3$GeTe$_2$ layer influences the exchange bias when interacting with other vdW antiferromagnets.

**Methods**

Few layered flakes of Fe$_3$GeTe$_2$ and CrPS$_4$ were exfoliated onto Si/SiO$_2$ (300 nm) substrates using standard mechanical exfoliation with scotch tape. The bulk crystals were characterized employing Raman spectroscopy (please refer to *Figure 1* in *Supporting Information* under *Section 1*). Suitable flakes of uniform thickness were located using an optical microscope. The located Fe$_3$GeTe$_2$ and CrPS$_4$ flakes were picked up and transferred in the order mentioned, by typical polymer based dry transfer method, to pre-patterned Hall contacts of Au/Cr (25 nm/5 nm) on Si/SiO$_2$ (300 nm) fabricated employing standard e-beam lithography. In order to investigate the impact of the surface oxide formed at the FM/AFM interface on EB, the surface of Fe$_3$GeTe$_2$ was exposed to an ambient atmosphere for 30 minutes before transferring the CrPS$_4$ layer resulting in the formation of an ultra-thin oxide layer, which is referred to as (O-FGT). To compare, another CrPS$_4$/ Fe$_3$GeTe$_2$ heterostructure with a pristine interface was fabricated the process of which was completely carried out inside the glovebox, containing an inert atmosphere (argon gas), and oxygen and moisture levels were kept below 0.5 ppm, to make sure there is no effect of any oxidation. The fabricated device with a pristine interface was then capped by hexagonal boron nitride (h-BN) flakes to prevent any chances of oxidation in air during sample manipulation. The surface oxide formation was confirmed using element-sensitive cross-sectional transmission electron microscopy measurements (please refer to *Figure 2* in *Supporting Information* under *Section 2* for details) of an exposed and annealed $Fe_3GeTe_2$ flake with a $CrPS_4$ flake stamped on top afterwards (*Figure S2a-h*). The thickness of the FM and AFM layers of the exposed heterostructure was determined to be $30 \pm 2\ nm$ and $100 \pm 2\ nm$, respectively, using atomic force microscope whereas it was found to be $33.3 \pm 11.2$ nm and $43.7 \pm 9.8$ nm for the capped device (please refer to *Figure S3* in *Supporting Information* under *Section 3*). Both devices with a pristine and exposed interfaces were then wire-bonded and immediately loaded onto a variable temperature insert (VTI) cryostat in which we can apply a magnetic field up to 12 T for anomalous Hall effect measurements. Since $CrPS_4$ is an insulating antiferromagnet, the current will only flow through the metallic FGT layer. The cross-sectional area of the current channel ($10\ \mu m\ x\ 5\ \mu m$) is $1.5 \times 10^{-13}$ $m^2$, and is used to calculate the current density, which is approximately $6.7 \times 10^6$ $Am^{-2}$ along a 30 nm $Fe_3GeTe_2$ flake for an applied current of $1\ \mu A$.

A $1\ \mu A$ current was passed through the $Fe_3GeTe_2$ flake along x-direction and the anomalous Hall voltage ($V_{xy}$) was measured across the transverse terminals along y-direction. The field was applied OOP orientation always for field-cooling, for field sweeping and for application of a *pre-set field\**. A Keithley 2400 source meter was used to flow current through the device, and Keithley 2182a nanovoltmeter was used to measure voltage.

*\*Pre-set field* ($H_{PF}$): The pre-set field, which is an out-of-plane field, is the field to which the device was exposed at a specific temperature just before starting the measurement of transverse

anomalous Hall voltage ($V_{xy}$). It is important to note that the pre-set field and field cooling are different because pre-set field doesn't involve any cooling process through the $T_N$ of the antiferromagnet. The mechanism of how a pre-set field determines the direction of pinning and hence the direction of loop shift is provided in *Figure S7* under *Supporting Information Section 7*.

**Abbreviations**

EB – Exchange Bias
$H_{EB}$ – Exchange Bias Field
$V_{xy}$ – Anomalous Hall Voltage
$H_{PF}$ – Pre-set Field
$T_B$ – Blocking Temperature
$T_N$ – Néel Temperature
O-FGT – Oxide layer formed on $Fe_3GeTe_2$ flakes

FIGURES

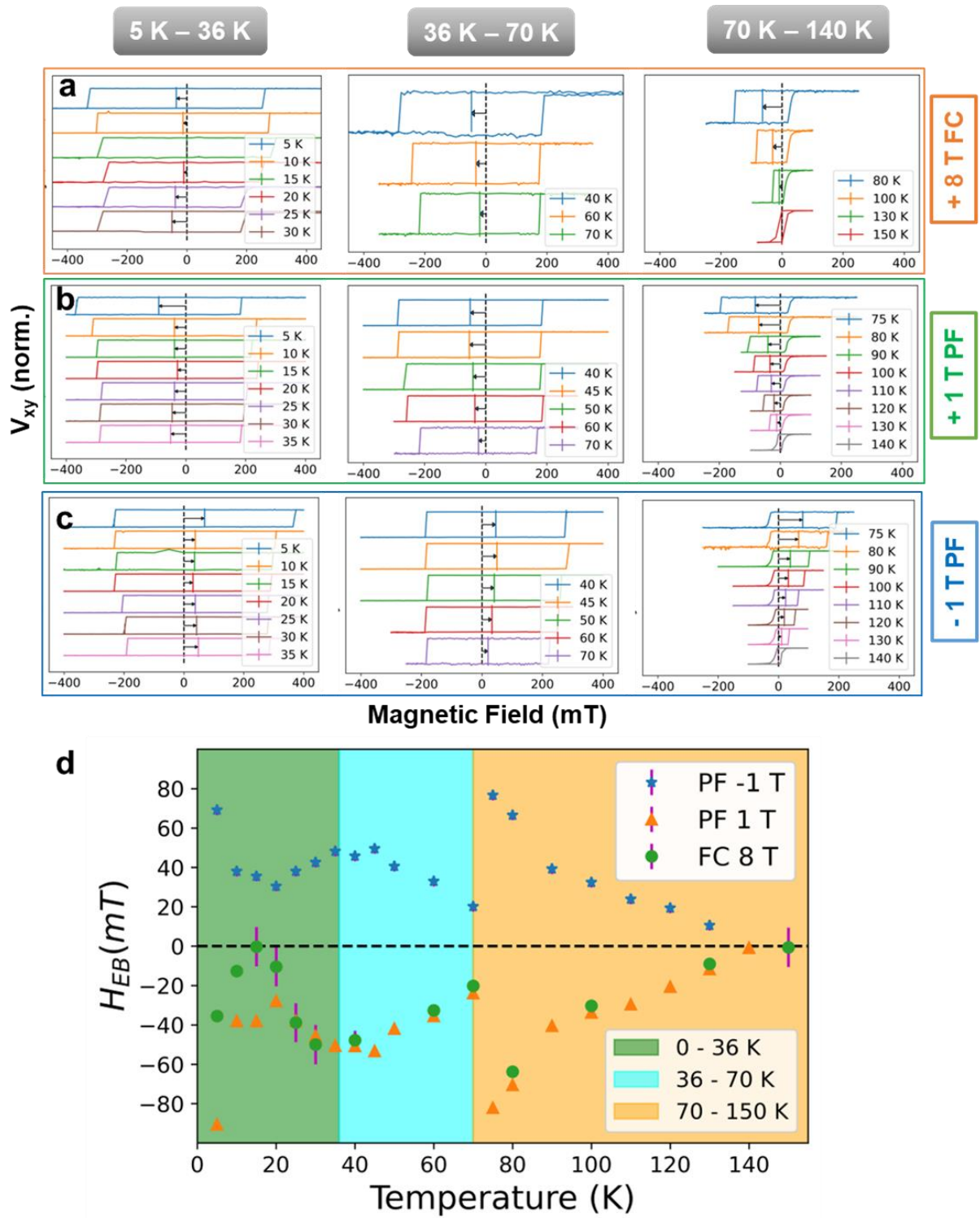

**Figure 1**. **EB due to surface oxide layers formed at the CrPS$_4$-Fe$_3$GeTe$_2$ interface:** Anomalous Hall voltage ($V_{xy}$) measured as a function of sweeping magnetic field for a CrPS$_4$/ (O-FGT) Fe$_3$GeTe$_2$ with **(a)** + 8 T field-cooling (FC), **(b)** +1 T pre-set field (PF), **(c)** -1 T pre-

set field, and **(d)** trend of exchange bias field ($H_{EB}$) calculated as a function of temperature for both +ve (orange) and -ve (blue) 1 T pre-set field experiments as well as for the +8 T field-cooled experiments (green).

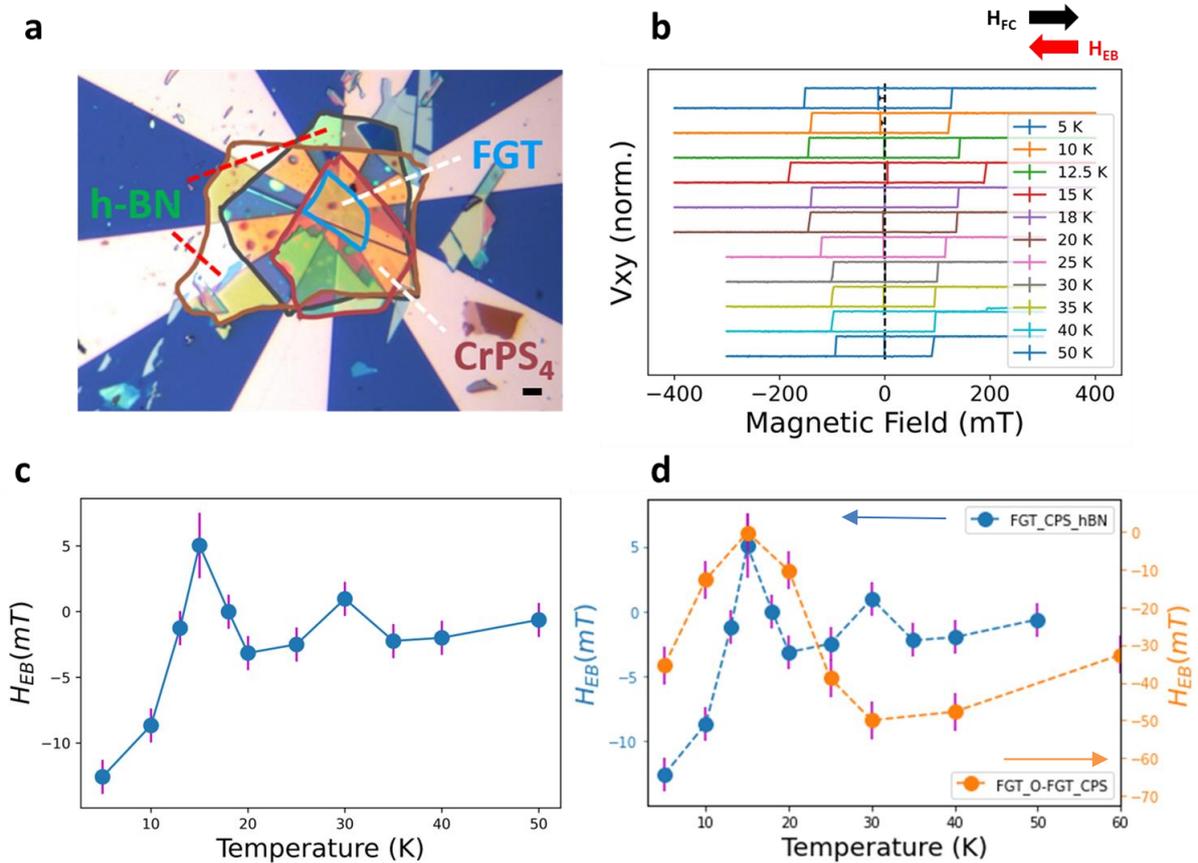

**Figure 2**. **EB due to a pristine CrPS$_4$/ Fe$_3$GeTe$_2$ interface:** **(a)** Optical micrograph of hBN/CrPS$_4$/ Fe$_3$GeTe$_2$ device (scalebar 10 μm), **(b)** Anomalous Hall voltage ($V_{xy}$) measured as a function of sweeping magnetic field for the capped device at various temperatures from 5 K to 50 K for + 8 T field-cooling (directions of $H_{FC}$ & $H_{EB}$ are indicated by black and red arrows respectively), **(c)** The trend of $H_{EB}$ as a function of temperature for + 8 T field-cooling, and **(d)** Comparison of $H_{EB}$ for a pristine and for an oxygen - exposed devices for + 8T field-cooling.

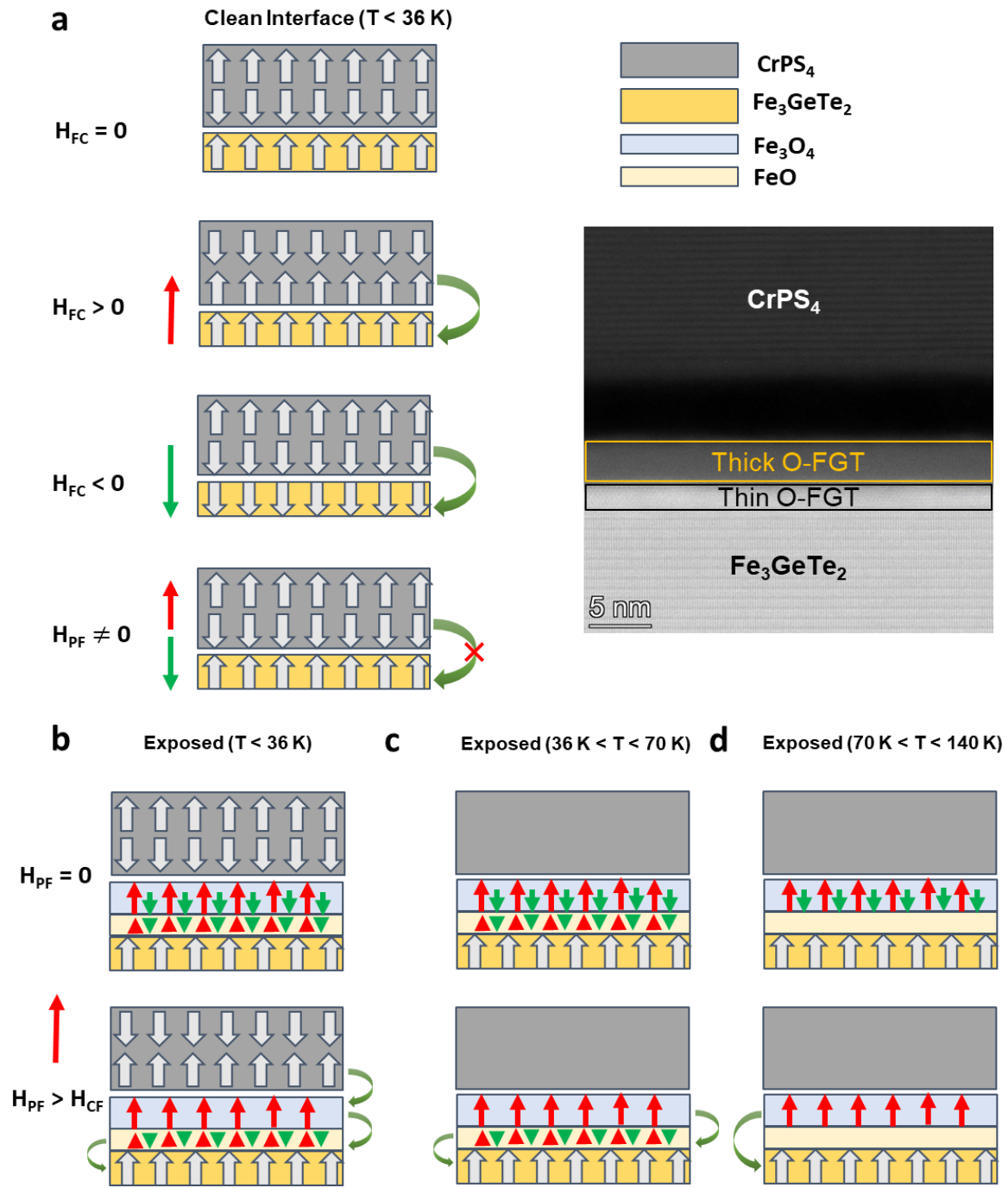

**Figure 3. Schematic Model:** **(a)** Exchange coupling in a pristine interface CrPS$_4$/ Fe$_3$GeTe$_2$ device after field-cooling (H$_{FC}$) with the cross-sectional STEM image of the CrPS$_4$/ (O-FGT) Fe$_3$GeTe$_2$ in the inset, and the mechanism of exchange coupling induced by a pre-set field (H$_{PF}$) at different temperature ranges **(b)** T < 36 K **(c)** 36 K < T < 70 K and **(d)** T > 70 K. The out-of-plane +ve and -ve field directions are indicated by the long red and green arrows, the exchange coupling is indicated by curved arrows.

**Author Contributions**

APB, AK and MK conceive the idea, design and carried out the experiment. TS and BL synthesized the $Fe_3GeTe_2$ bulk crystals used for the experiments. ZCL and JY synthesized $CrPS_4$ single crystals. SF and HW performed Raman measurements. TD, JV, AKV and RDB carried out the STEM measurements. AS and UN helped in interpretation of the results. APB, AK contributed equally. All the authors contributed to manuscript writing.

**Acknowledgements**

We acknowledge funding from Alexander von Humboldt Foundation for Humboldt Postdoctoral Fellowship (Grant number: Ref 3.5-IND-1216986-HFST-P), EU Marie-Curie Postdoctoral Fellowship ExBiaVdW (Grand Id: 101068014), Deutsche Forschungsgemeinschaft (DFG, German Research Foundation) – Spin + X TRR 173–268565370 (Projects No. A01 and No. B02), DFG Project No. 358671374, Graduate School of Excellence Materials Science in Mainz (MAINZ) GSC 266, the MaHoJeRo (DAAD Spintronics network, Projects No. 57334897 and No. 57524834), the Research Council of Norway (Centre for Quantum Spintronics - QuSpin No. 262633), the National Natural Science Foundation of China (Nos. 12241401 and 11975035), The European Union's Horizon 2020 Research and Innovation Programme under grant agreement 856538 (project "3D MAGIC"). S.F. acknowledges the China Scholarship Council for financial support.

**Supporting Information**

The supporting information encompasses material morphology and structural characterization measurements conducted using Raman Spectroscopy, Transmission Electron Microscopy, x-ray photoelectron spectroscopy, and atomic force microscope techniques (Sections 1 to 5). Additionally, further transport measurements are included as supporting evidence for the observed phenomena discussed in the manuscript (Section 6 and sections 8 to 11). A schematic diagram illustrating how a pre-set field induces exchange bias is provided in section 7.

**References**

1. Blachowicz, T.; Ehrmann, A. Exchange Bias in Thin Films—An Update. *Coatings* **2021**, *11*, 122.
2. Coehoorn, R. Giant magnetoresistance and magnetic interactions in exchange-biased spin-valves. In *Handbook of Magnetic Materials*, Vol. 15; Elsevier, 2003; pp 1-197.
3. Schuller, I. K.; Morales, R.; Batlle, X.; Nowak, U.; Güntherodt, G. Role of the antiferromagnetic bulk spins in exchange bias. *Journal of Magnetism and Magnetic Materials* **2016**, *416*, 2-9.
4. Nogués, J.; Schuller, I. K. Exchange bias. *Journal of Magnetism and Magnetic Materials* **1999**, *192*, 203-232.
5. Fernando, G. W. Chapter 4 - Magnetic Anisotropy in Transition Metal Systems. In *Handbook of Metal Physics*, Fernando, G. W. Ed.; Vol. 4; Elsevier, 2008; pp 89-110.
6. Gong, C.; Zhang, X. Two-dimensional magnetic crystals and emergent heterostructure devices. *Science* **2019**, *363*, eaav4450.
7. Phan, M.-H.; Kalappattil, V.; Jimenez, V. O.; Thi Hai Pham, Y.; Mudiyanselage, N. W. Y. A. Y.; Detellem, D.; Hung, C.-M.; Chanda, A.; Eggers, T. Exchange bias and interface-related effects in two-dimensional van der Waals magnetic heterostructures: Open questions and perspectives. *Journal of Alloys and Compounds* **2023**, *937*, 168375.
8. Zhu, R.; Zhang, W.; Shen, W.; Wong, P. K. J.; Wang, Q.; Liang, Q.; Tian, Z.; Zhai, Y.; Qiu, C.-w.; Wee, A. T. S. Exchange Bias in van der Waals CrCl3/Fe3GeTe2 Heterostructures. *Nano Letters* **2020**, *20*, 5030-5035.
9. Dai, H.; Cheng, H.; Cai, M.; Hao, Q.; Xing, Y.; Chen, H.; Chen, X.; Wang, X.; Han, J.-B. Enhancement of the Coercive Field and Exchange Bias Effect in Fe3GeTe2/MnPX3 (X = S and Se) van der Waals Heterostructures. *ACS Applied Materials & Interfaces* **2021**, *13*, 24314-24320.

10. Albarakati, S.; Xie, W. Q.; Tan, C.; Zheng, G.; Algarni, M.; Li, J.; Partridge, J.; Spencer, M. J. S.; Farrar, L.; Xiong, Y.; et al. Electric Control of Exchange Bias Effect in FePS3–Fe5GeTe2 van der Waals Heterostructures. *Nano Letters* **2022**, *22*, 6166-6172.

11. Ying, Z.; Chen, B.; Li, C.; Wei, B.; Dai, Z.; Guo, F.; Pan, D.; Zhang, H.; Wu, D.; Wang, X.; et al. Large Exchange Bias Effect and Coverage-Dependent Interfacial Coupling in CrI3/MnBi2Te4 van der Waals Heterostructures. *Nano Letters* **2023**, *23*, 765-771.

12. Xu, X.; Yang, S.; Wang, H.; Guzman, R.; Gao, Y.; Zhu, Y.; Peng, Y.; Zang, Z.; Xi, M.; Tian, S.; et al. Ferromagnetic-antiferromagnetic coexisting ground state and exchange bias effects in MnBi4Te7 and MnBi6Te10. *Nature Communications* **2022**, *13*, 7646.

13. Bo, X.; Li, F.; Yin, X.; Chen, Y.; Wan, X.; Pu, Y. Magnetic structure and exchange interactions of the van der Waals CrPS 4 monolayer under strain: A first-principles study. *Physical Review B* **2023**, *108*, 024405.

14. Calder, S.; Haglund, A. V.; Liu, Y.; Pajerowski, D. M.; Cao, H. B.; Williams, T. J.; Garlea, V. O.; Mandrus, D. Magnetic structure and exchange interactions in the layered semiconductor CrPS$_4$. *Physical Review B* **2020**, *102*, 024408.

15. Peng, Y.; Ding, S.; Cheng, M.; Hu, Q.; Yang, J.; Wang, F.; Xue, M.; Liu, Z.; Lin, Z.; Avdeev, M.; Hou, Y.; Yang, W.; Zheng, Y.; Yang, J. Magnetic structure and metamagnetic transitions in the van der Waals antiferromagnet CrPS$_4$. *Advanced Materials* **2020**, *32*, 2001200.

16. Fei, Z.; Huang, B.; Malinowski, P.; Wang, W.; Song, T.; Sanchez, J.; Yao, W.; Xiao, D.; Zhu, X.; May, A. F.; et al. Two-dimensional itinerant ferromagnetism in atomically thin Fe3GeTe2. *Nature Materials* **2018**, *17*, 778-782.

17. Gweon, H. K.; Lee, S. Y.; Kwon, H. Y.; Jeong, J.; Chang, H. J.; Kim, K.-W.; Qiu, Z. Q.; Ryu, H.; Jang, C.; Choi, J. W. Exchange Bias in Weakly Interlayer-Coupled van der Waals Magnet Fe3GeTe2. *Nano Letters* **2021**, *21*, 1672-1678.

18. Kim, D. S.; Kee, J. Y.; Lee, J.-E.; Liu, Y.; Kim, Y.; Kim, N.; Hwang, C.; Kim, W.; Petrovic, C.; Lee, D. R.; et al. Surface oxidation in a van der Waals ferromagnet Fe3-xGeTe2. *Current Applied Physics* **2021**, *30*, 40-45.

19. Li, Y.; Hu, X.; Fereidouni, A.; Basnet, R.; Pandey, K.; Wen, J.; Liu, Y.; Zheng, H.; Churchill, H. O. H.; Hu, J.; et al. Visualizing the Effect of Oxidation on Magnetic Domain Behavior of Nanoscale Fe3GeTe2 for Applications in Spintronics. *ACS Applied Nano Materials* **2023**, *6*, 4390-4397.

20. Zhang, T.; Zhang, Y.; Huang, M.; Li, B.; Sun, Y.; Qu, Z.; Duan, X.; Jiang, C.; Yang, S. Tuning the Exchange Bias Effect in 2D van der Waals Ferro-/Antiferromagnetic Fe3GeTe2/CrOCl Heterostructures. *Advanced Science* **2022**, *9*, 2105483.

21. Ma, S.; Li, G.; Li, Z.; Zhang, Y.; Lu, H.; Gao, Z.; Wu, J.; Long, G.; Huang, Y. 2D magnetic semiconductor Fe3GeTe2 with few and single layers with a greatly enhanced intrinsic exchange bias by liquid-phase exfoliation. *ACS Nano* **2022**, *16*, 19439-19450.

22. Ding, S.; Peng, Y.; Xue, M.; Liu, Z.; Liang, Z.; Yang, W.; Sun, Y.; Zhao, J.; Wang, C.; Liu, S.; et al. Magnetic phase diagram of CrPS4 and its exchange interaction in contact with NiFe. *Journal of Physics: Condensed Matter* **2020**, *32*, 405804.

23. Och, M.; Martin, M.-B.; Dlubak, B.; Seneor, P.; Mattevi, C. Synthesis of emerging 2D layered magnetic materials. *Nanoscale* **2021**, *13*, 2157-2180.

24. Kong, X.; Berlijn, T.; Liang, L. Thickness and Spin Dependence of Raman Modes in Magnetic Layered Fe3GeTe2. *Advanced Electronic Materials* **2021**, *7*, 2001159.

25. Wu, H.; Chen, H. Probing the properties of lattice vibrations and surface electronic states in magnetic semiconductor CrPS4. *RSC Advances* **2019**, *9*, 30655-30658.

26. Mauri, D.; Siegmann, H. C.; Bagus, P. S.; Kay, E. Simple model for thin ferromagnetic films exchange coupled to an antiferromagnetic substrate. *Journal of Applied Physics* **1987**, *62*, 3047–3049.

27. Radu, F.; Abrudan, R.; Radu, I.; Schmitz, D.; Zabel, H. Perpendicular exchange bias in ferrimagnetic spin valves. *Nature Communications* **2012**, *3*, 715.

28. Dho, J. Magnetic-field-induced switchable exchange bias in NiFe film on (110) Fe3O4 with a strong uniaxial magnetic anisotropy. *Applied Physics Letters* **2005**, *106*, 202405.

(29) Kozioł-Rachwał, A.; Ślęzak, T.; Nozaki, T.; Yuasa, S.; Korecki, J. Growth and magnetic properties of ultrathin epitaxial FeO films and Fe/FeO bilayers on MgO(001). *Applied Physics Letters* **2016**, *108*, 041606.

TOC

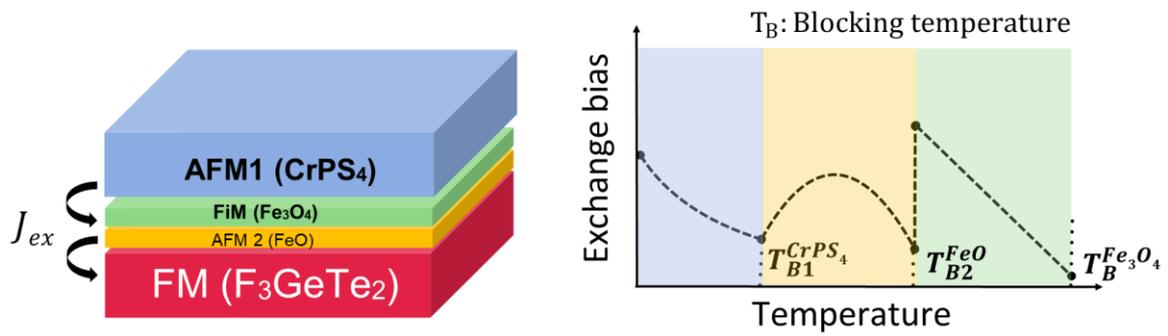

**Summary of Research:** A non-monotonic and complex trend of the exchange bias as a function of the temperature due to the surface oxide formation on $Fe_3GeTe_2$ at the interface of the $CrPS_4$/$Fe_3GeTe_2$ AFM/FM heterostructure.